\documentclass[amssymb,amsmath,aps]{revtex4}   

\usepackage[dvipdfm]{graphicx}
\usepackage{latexsym,mathrsfs}
\usepackage{dcolumn}
\usepackage{bm}
\usepackage{graphicx}
\usepackage{txfonts}
\usepackage{natbib}
\usepackage{amsmath}    
\usepackage{graphicx}   

\begin{document}

\title{Hierarchical Bass model}

\author{Tohru Tashiro}
\email{tashiro@cosmos.phys.ocha.ac.jp}   
\affiliation{Department of Physics, Ochanomizu University, 2-1-1 Ohtsuka, Bunkyo, Tokyo 112-8610, Japan}
\date{\today}

\begin{abstract}
We propose a new model about diffusion of a product which includes a memory of how many {\em adopters} or advertisements a {\em non-adopter} met, where (non-)adopters mean people (not)  possessing the product. This effect is lacking in the Bass model. As an application, we utilize the model to fit the iPod sales data, and so the better agreement is obtained than the Bass model.
\end{abstract}

\maketitle

\section{Introduction}

It is so interesting how fashion diffuses in a society.
It has been fascinating scientists and scholars how a new product or innovation spreads in a society.
``No other field of behavior science research represents more effort by more scholars in more disciplines in more nations.''  \cite{Rogers03}
We shall employ a penetration rate of a product in studying a fashion diffusion scientifically because it is an objective and available statistical data.
Generally, a change of the penetration rate in time is slow at first, and then it peaks, followed again by slow change.
The logistic model (LM) can describe this change in time. LM is represented by
\begin{equation}
\frac{{\rm d}P(t)}{{\rm d}t} = \frac{a}{N}P(t)\left\{N-P(t)\right\} \ ,
\label{eq:logi}
\end{equation}
where $P(t)$ is the number of those possessing the product at $t$ and $N$ is the total population.
In Ref.~\cite{Tashiro13}, we made it clear that the communication between individuals is imitation; people who do not possess (non-adopters) purchase the product quickly after they met people who already possess it (adopters).
However, the communication is unnatural, because there are not only trend-conscious people, but also cautious people in our society.
We considered that a memory of how many adopters a non-adopter met is essential for him/her to decide to purchase the product.
Accordingly, we proposed a model including this effect, which is the {\em hierarchical logistic model } (HLM) \cite{Tashiro13}.

Another essential element for fashion diffusion is an advertisement.
The Bass model (BM) \cite{Bass69} includes the element, which is represented by
\begin{equation}
\frac{{\rm d}P(t)}{{\rm d}t} = \left\{b+\frac{a}{N}P(t)\right\}\left\{N-P(t)\right\} \ ,
\label{eq:Bass}
\end{equation}
which coincides with LM by setting $b=0$.
This model has been utilized for analyzing the penetration of various kinds of new products and technical innovations: Mobile phones \cite{Dekimpe98,Michalakelis08,Chuetal09,Limetal12}, Internet \cite{Fornerino03,Turk12}, wireless communication \cite{Sundqvist05}, photovoltaic systems \cite{Guidolin10}, hydrogen fuel cell vehicles \cite{Park11} and so on.

Indeed, BM does not  take account of the memory effect we mentioned above.
So, in this study, we shall propose a model including it, which is the {\em hierarchical Bass model} (HBM).

This paper is organized as follows. In Sec.~\ref{BM}, BM is yielded by our original way like Ref.~\cite{Tashiro13}.
Then, in Sec.~\ref{HBM}, we construct HBM by including the memory effect in BM.
Lastly, HBM is applied to fitting the iPod sales data in Sec.~\ref{fitting}, and then the better agreement is obtained than BM.

\section{Bass model}
\label{BM}

In this section, we derive BM in the same way as Ref.~\cite{Tashiro13}.
Let us put $N$ random walkers on $n\times n \ (> N)$ lattices. They move to one of the nearest lattices with the same provability at each discretized time step. Furthermore, we shall apply the following rules for the random walkers:
i) If non-adopters meet adopters, they start to adopt a product quickly,
ii) there are no interactions among (non-)adopters, and 
iii) adopters do not part with it. 
Here, we define sharing a lattice with someone as meeting the person.
By these rules, we can derive LM \cite{Tashiro13}.
In order to derive BM, we need to consider another influence on non-adopters from other than adopters, that is an advertisement like a billboard. 
We assume that the number of advertisements does not change in time. In addition, we apply the forth rule:
iv)  Non-adopters who met advertisements start to adopt the product quickly.
As a final assumption, we set the density of people to be so low that three or more people cannot share a lattice at the same step, as if three or more body collisions are ignored in a rarefied gas. Therefore, it is natural to consider that a probability of meeting adopters, non-adopters, or advertisements is proportional to each number of them.

Let us set the number of adopters and non-adopters at the $i$th step as $P_i$ and $Q_i$, respectively. Indeed, $P_i+Q_i=N$. Moreover, we set the number of advertisements $B \ (<n^2)$. Then, the probability of meeting adopters, non-adopters and advertisements can be represented by $P_i/n^2$, $Q_i/n^2$ and $B/n^2$, respectively.
According to the first and the forth rule, $(B/n^2)\times Q_i+(P_i/n^2)\times Q_i$ people of non-adopters becomes adopters at the next step, and then, the following recursion formulae can be obtained.
\begin{align}
P_{i+1} &= P_i  + \frac{B}{n^2}Q_i + \frac{P_i}{n^2}Q_i \\
Q_{i+1} &= Q_i -  \frac{B}{n^2}Q_i - \frac{P_i}{n^2}Q_i  \ .
\end{align}

We shall define the number of them at $t$ as $P(t) = P(i\cdot\Delta t) \equiv P_i$ and $Q(t) = Q(i\cdot\Delta t) \equiv Q_i$.
Here, we take the limits as $\Delta t\rightarrow 0$ and $n\rightarrow\infty$ with $n^2\Delta t$ fixed. 
By setting the fixed value as 
\begin{equation}
\frac{a}{N} = \lim_{\stackrel{\scriptstyle \Delta t\rightarrow 0}{n\rightarrow\infty}}\frac{1}{n^2\Delta t} \ ,
\label{firsta}
\end{equation}
defining $Ba/N$ as 
\begin{equation}
\lim_{\stackrel{\scriptstyle \Delta t\rightarrow 0}{n\rightarrow\infty}}\frac{B}{n^2\Delta t} = \frac{Ba}{N} \equiv b\ ,
\label{firstb}
\end{equation}
and using $P(t)+Q(t)=N$, the following differential equation is derived:
\begin{equation}
\frac{{\rm d}P(t)}{{\rm d}t} = \left\{b+\frac{a}{N}P(t)\right\}\left\{N-P(t)\right\} , 
\end{equation}
which describes BM. If we use the penetration rate $p(t)\equiv P(t)/N$, the above differential equation becomes
 \begin{equation}
\frac{{\rm d}p(t)}{{\rm d}t} = \left\{b+ap(t)\right\}\left\{1-p(t)\right\} \ .
\end{equation}
Note that $a>0$ and $b\ge0$ by the definitions Eqs.~(\ref{firsta}) and (\ref{firstb}).

\section{Hierarchical Bass model}
\label{HBM}

In the previous section, it is unveiled that BM reflects quick influences on non-adopters from adopters and advertisements. However, it is unnatural because all the people in our society are not easily influenced by others and advertisements. 
We consider that a memory of how many adopters or advertisements a non-adopter met is important for his/her decision to purchase the product.
Therefore, as in Ref.~\cite{Tashiro13}, we shall include the memory in BM by the following way:
We set the number of non-adopters starting to adopt the product after they meet $\mu$ adopters or advertisements at $i$th step as $Q^\mu_i$, in which we call $\mu$ as {\em remaining adopters and advertisements number} (RAAN).
Indeed, if a non-adopter, whose RAAN is $\mu$, meet one of adopters or advertisements, his/her RAAN becomes $\mu-1$
at the next step. We do not alter other rules.

If the maximum of RAAN is $m$, the recursion formulae change into
\begin{align}
P_{i+1}          &= P_i + \left(\frac{B}{n^2}+\frac{P_i}{n^2}\right)Q_i^{1}  \ , \\
Q_{i+1}^{1}  &= Q_i^{1} -\left(\frac{B}{n^2}+\frac{P_i}{n^2}\right)Q_i^{1}
+ \left(\frac{B}{n^2}+\frac{P_i}{n^2}\right)Q_i^{2}  \ ,  \\
                      &\vdots  \nonumber \\
Q_{i+1}^{m} &= Q_i^{m} -\left(\frac{B}{n^2}+\frac{P_i}{n^2}\right)Q_i^{m} \ .
\end{align}
These become the following differential equations by the previous continuation of space and time.
\begin{align}
\frac{{\rm d}P(t)}{{\rm d}t} &= \left\{b+\frac{a}{N}P(t)\right\}Q^1(t) \\
\frac{{\rm d}Q^1(t)}{{\rm d}t} &= -\left\{b+\frac{a}{N}P(t)\right\}Q^1(t) + \left\{b+\frac{a}{N}P(t)\right\}Q^2(t) \\
                                                   &\vdots  \nonumber \\
\frac{{\rm d}Q^m(t)}{{\rm d}t} &= -\left\{b+\frac{a}{N}P(t)\right\}Q^m(t) \label{eq:pq3}
\end{align}
where $Q^\mu(t) = Q^\mu(i\cdot\Delta t) \equiv Q_i^\mu$.
We shall name this the {\em hierarchical Bass model} (HBM).
Indeed, $P(t) + Q^1(t) + \cdots + Q^{m}(t) = N$ is always conserved.

We can solve Eq.~(\ref{eq:pq3}) easily: The solution is 
\begin{equation}
  Q^m(t) = Q^m(0)\exp\left[-bt-\frac{a}{N}\int_{0}^{t}{\rm d}t'P(t')\right] \ . 
\end{equation}
If $Q^m(0)=0$, $Q^m(t)$ is always 0. Then, the contribution of $Q^m$ into the differential equation of $Q^{m-1}$ dismisses, and so $Q^{m-1}$ can be calculated similarly. Therefore, if $Q^{m}(0)=Q^{m-1}(0)=\cdots=Q^2(0)=0$, $Q^{m}(t)=Q^{m-1}(t)=\cdots=Q^2(t)=0$, which means BM is recovered. Namely, HBM includes BM.

If we use ratios of adopters and non-adopters to the total number $N$, HBM becomes
\begin{align}
\frac{{\rm d}p(t)}{{\rm d}t} &= \left\{b+ap(t)\right\}q^1(t) \label{eq:hbmF} \\
\frac{{\rm d}q^1(t)}{{\rm d}t} &= -  \left\{b+ap(t)\right\}q^1(t) + \left\{b+ap(t)\right\}q^2(t) \\
                                                   & \vdots  \nonumber \\
\frac{{\rm d}q^m(t)}{{\rm d}t} &= -  \left\{b+ap(t)\right\}q^m(t) \label{eq:hbmL}
\end{align}
where $q^\mu(t)\equiv Q^\mu(t)/N$.

In order to examine the solution of HBM, we devide the both sides of Eqs.~from (\ref{eq:hbmF}) to  (\ref{eq:hbmL}) by $a$, and then, define the scaled time by $1/a$ as $\tau$. Therefore, the solutions of these nondimensionalized equations are characterized by only one parameter, $b/a$. However, a combination of initial condition increases as $m$ becomes larger. So, we shall confine the discussion to the following initial condition:
\begin{equation}
p(0) = 0 \ , \ \ q^{\mu}(0) = \left\{
\begin{array}{cl}
1 & (\mu = m) \\
0 & (\mbox{otherwise})
\end{array}
\right. \ . \label{eq:IC}
\end{equation}

\begin{figure}[h]
    \includegraphics[scale=1.05]{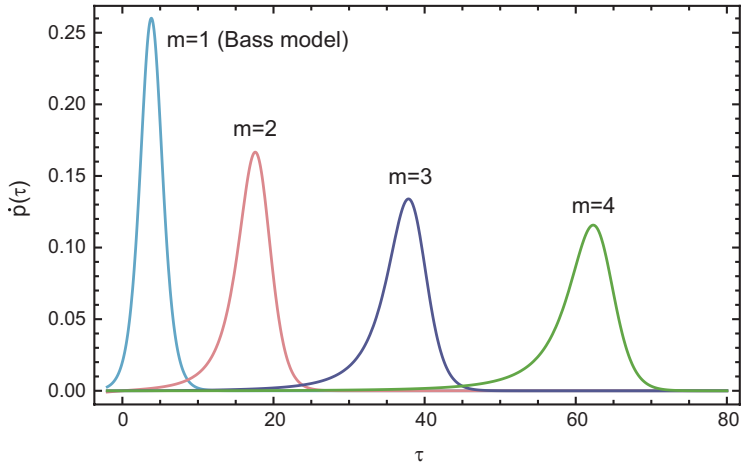}\hspace{2pc}%
    \begin{minipage}[b]{14pc}\caption{\label{Fig:ppra}(color online) The derivative of $p(\tau)$, $\dot{p}(\tau)$, with $m=1 \ (\mbox{Bass model})$, 2,  3 and 4 vs. the saceled time $\tau$ for $b/a=0.02$.}
  \end{minipage}
\end{figure}

One can easily suppose that a growth rate of $p(\tau)$ becomes slower than BM because of the hierarchal structure of HBM. 
Figure~\ref{Fig:ppra} showing the derivative of $p(\tau)$, $\dot{p}(\tau)$, for several $m$ reflects this fact. 
In addition, we can see that a symmetry of the growth rate at the peak breaks down for $m\ge 2$ from this figure.
As well know, the growth rate of BM is symmetric regardless of the initial condition (See Appendix).
For HBM, however, this property does not hold. 

\begin{figure}[h]
    \includegraphics[scale=1.05]{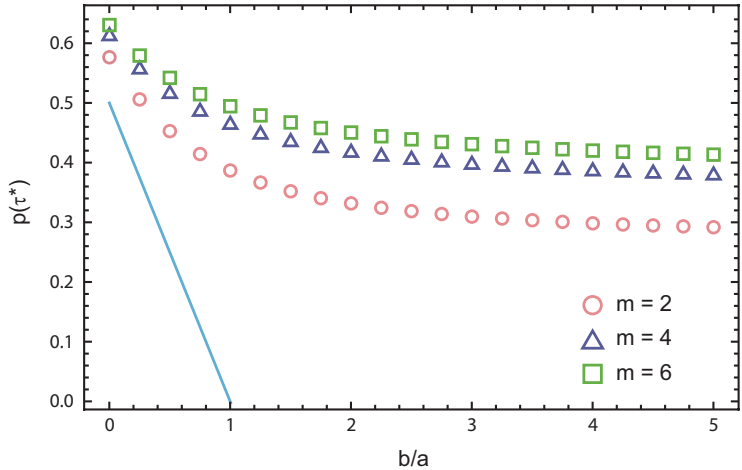}\hspace{2pc}%
    \begin{minipage}[b]{14pc}\caption{\label{Fig:pforpeak}(color online) $p(\tau^*)$ as a function of $b/a$ for $m=2$, 4 and 6 where $\tau^*$ is a time at which the derivative of $p(\tau)$ peaks. The line means a relation between $p(\tau^*)$ and $b/a$, $p(\tau^*) = (1-b/a)/2$, for $m=1 \ (\mbox{Bass model})$.}
  \end{minipage}
\end{figure}

Here, let us pay attention to a relation between $p(\tau^*)$ and $b/a$ where $\tau^*$ is a time at which $\dot{p}(\tau)$ peaks, that is $\ddot{p}(\tau^*)=0$. This relation for BM is $p(\tau^*) = (1-b/a)/2$ (see Appendix), which means that $p(\tau^*)$ does not extend beyond $50\%$ and $\dot{p}(\tau)$ has no peaks if $b/a\ge1$ regardless of the initial condition: ${p}(\tau)$ has no inflection points. This constraint, ${p}(\tau^*)\le0.5$, is one of the conceptional limitations pointed out by Ref.~\cite{Easing83} where they overcome this limitation by introducing a new parameter $\delta$.
On the other hand,  one can see that the relation for HBM becomes nonlinear and $p(\tau^*)$ can exceed $50\%$ from Fig.~\ref{Fig:pforpeak}.

\section{fitting iPod sales data}
\label{fitting}
\begin{figure}[h]
    \includegraphics[scale=0.65]{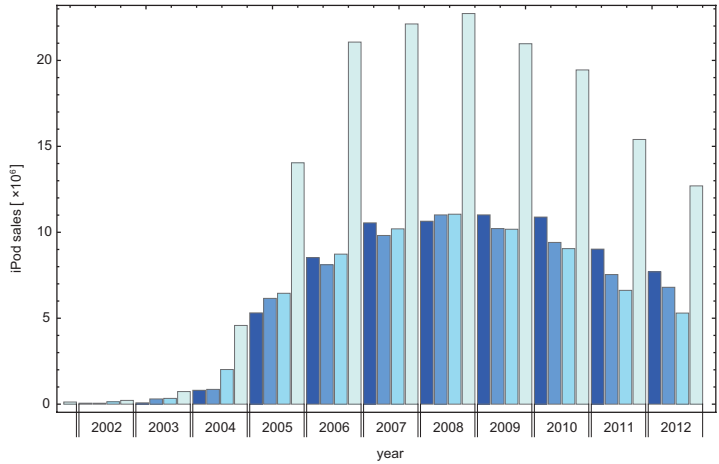}\hspace{2pc}%
    \begin{minipage}[b]{14pc}\caption{\label{Fig:iPod}(color online) iPod sales data obtained from  Apple Inc.'s official homepage, http://www.apple.com/. Each year includes four quarters, February, May, August and November. The same (color) gray-level means the same quarter. The first bar is the data on November, 2001.}
  \end{minipage}
\end{figure}

Let us apply HBM to analyze a sales data of a product in this section. For it, we shall employ the iPod sales as in  Ref.~\cite{Tashiro13}.
We can obtain the data from the official homepage of Apple Inc. which has been creating and marketing the iPod: http://www.apple.com/.
As of September 2013, the data from the end of 2001 to the end of 2012 has been reported quarterly, which we show in Fig.~\ref{Fig:iPod}. We can see that there are eight peaks in the figure.
According to the rules in Sec.~\ref{BM} and \ref{HBM}, BM and HBM do not suppose that adopters purchase the new product.
Because Apple Inc. has marketed many types of iPod since November 2001, however, many adopters must purchase several iPods until now.
Therefore, we shall utilize the data from November 2001 to May 2006, for the interval includes only one peak of sales and has a relatively small effect of  the repurchase. 

Setting November 2001 as the origin of the time axis, we construct the cumulative sales,  and then, we fit the sales data by HBM, BM and LM.
In doing so, if we minimize the residual sum of squares or the sum of the absolute value of error (SAE), the fit between the model and the data is good for high values of sales but not for low ones.
On the other hand, if we minimize the sum of the absolute value of relative error (SARE), the fit for low values becomes better. We consider that this is because the maximum of iPod sales data is 3 digits different than the minimum.
Thus, in order to resolve the problem, we shall minimize a product of SAE and SARE. The parameters minimizing this product are show in Tab.~\ref{Tab:para}, together with ones of LM.

From the table, we can see that the product and SARE diminish with increase of $m$: SARE with $m=4$ reduces to nearly half that with $m=1$. In other words, the average relative error for HBM with $m=4$ is about as half as that for BM. 
For reference, the coefficient of determination $R^2$ is shown in Tab.~\ref{Tab:para}. The closer the value is to 1, the smaller the residual sum of squares is.
It can be seen that $R^2$s of HBM are closer to 1. So, we can argue that the residual and the relative error between HBM and the data are small and HBM can approximate the whole sales data.

Results of fitting the data by HLM are also shown in Tab.~\ref{Tab:para} by numbers in parentheses. From these, we can say that HBM considering the effect of advertisements can approximate the sales data better than HLM. However, it is not a reason that there is an extra fit parameter $b$ in HBM, because the degrees of agreement of LM and BM are comparable.

\begin{table}
\caption{\label{Tab:para}Parameters of HBM and LM minimizing the product of SARE and SAE, SAREs and $R^2$s by them. Numbers in parentheses mean results of fitting the data by HLM. All values are rounded to a three-digit number, and so the sum of $p(0)$ and $p^\mu(0)$ is not equal to one.}
\begin{center}
\begin{tabular*}{1.0\textwidth}{p{0.13\textwidth}p{0.12\textwidth}p{0.15\textwidth}p{0.16\textwidth}p{0.16\textwidth}p{0.16\textwidth}}
\hline
\hline
                               &                                     &\multicolumn{4}{c}{HBM} \\
                               &\raisebox{2ex}{LM}   &$m=1$ \ (BM)              &$m=2$                        &$m=3$         &$m=4$  \\
\hline
$a$                         & $1.42$                         & $1.42$                        & $4.22$                       & $4.25$       & $4.50$       \\
$b$                         & $-$            & $2.19\times10^{-10}$& $7.16\times10^{-4}$& $6.93\times10^{-4}$&$7.10\times10^{-4}$  \\
$N \ [\times10^6]$& $1.28\times10^5$      & $1.07\times10^5$      & $65.7$                       & $65.6$       & $66.4$      \\
$p(0)$                     & $9.26\times10^{-7}$ & $1.11\times 10^{-6}$& $0.00189$                 & $0.00193$& $0.00188$ \\
$q^1(0)$                 & $1.00$                        & $1.00$                        & $0.309$                    & $0.305$     & $0.290$     \\
$q^2(0)$                 & $-$                             & $-$                              & $0.689$                    & $0.686$     & $0.640$     \\
$q^3(0)$                 & $-$                             & $-$                              & $-$                           & $0.00739$ & $0.00800$ \\
$q^4(0)$                 & $-$                             & $-$                              & $-$                           & $-$             & $0.0600$  \\
\hline
product                   &$62.5$                        & $62.5$               & $8.73$\ ($11.0$)     & $8.67$\ ($8.91$)&$8.33$\ ($8.35$) \\
SARE                       & $1.84$                       & $1.84$               & $0.999$\ ($1.12$)   & $0.989$\ ($1.02$)&$0.984$\ ($1.02$) \\
$R^2$                     & $0.959$                     & $0.959$             & $0.998$\ ($0.997$)  & $0.998$\ ($0.998$)&$0.998$\ ($0.998$) \\
\hline
\hline
\end{tabular*}
\end{center}
\end{table}

From the fit parameters of HBM with $m=4$, we can guess the following facts: The market size producing the first peak of iPod sales is about 66 million people, the ratio of the trend-conscious people, $q^1(0)$, is about 29\%, the ratio of the cautious people, $q^2(0)$, is about 64\%, and the ratio of more cautious people, $q^3(0)+q^4(0)$, is about 7\%.
 
\begin{figure}[h]
  \begin{center}
    \begin{minipage}{14pc}
      \hspace{-3pc}
      \includegraphics[scale=0.8]{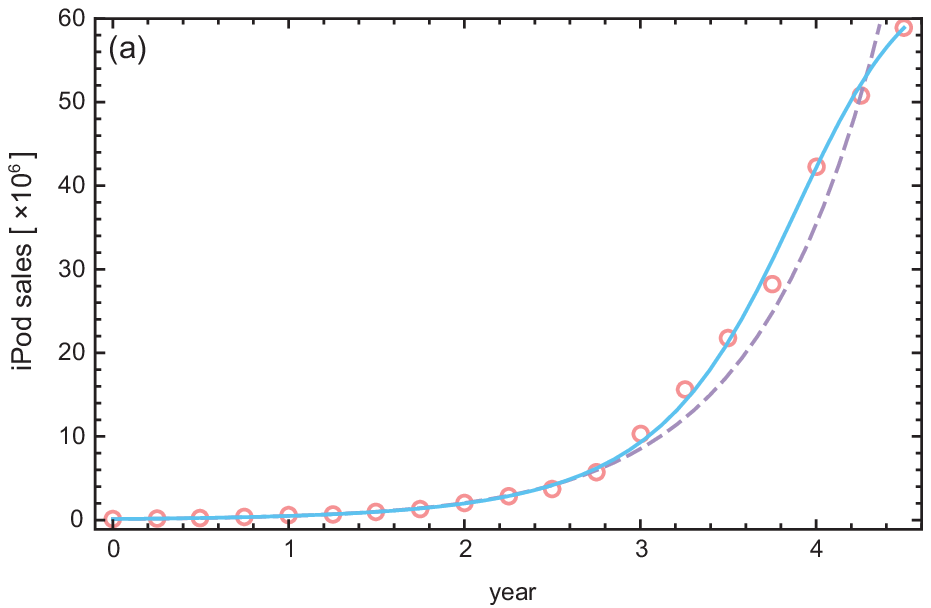}
    \end{minipage}
    \hspace{5pc}%
    \begin{minipage}{14pc}
      \hspace{-2pc}
      \includegraphics[scale=0.8]{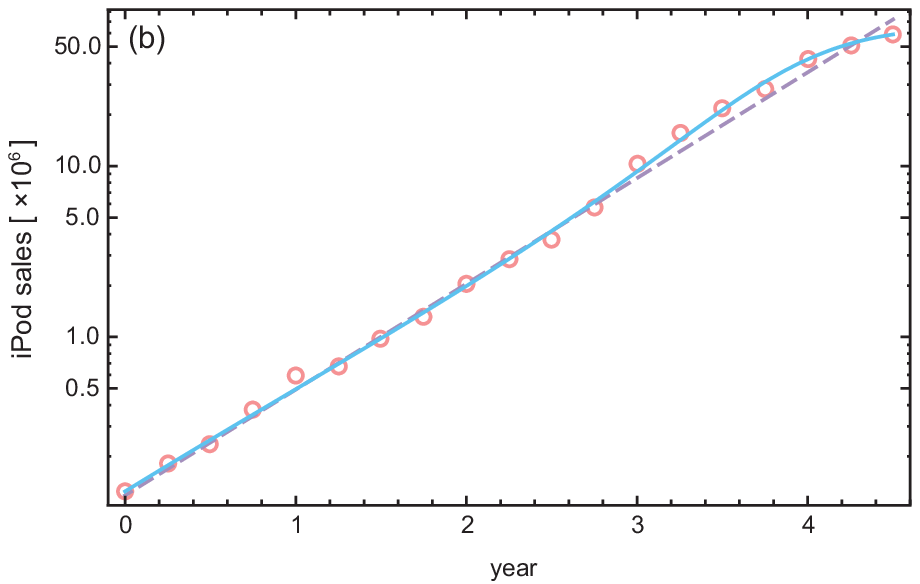}
    \end{minipage} 
    \caption{\label{Fig:m1and4}(color online) Cumulative iPod sales data represented by circles and fitting curves. The (purple) dashed and the (light blue) full curves are BM and HMB with $m=4$, respectively. (b) shows the same plot as (a) in logarithmic scale.}
  \end{center}
\end{figure}

In order to compare the degrees of agreement of BM and HBM with $m=4$, we plot the sales data with them in Fig.~\ref{Fig:m1and4}.
The circles are the cumulative iPod sales and the (purple) dashed and the (light blue) full curves represent BM and HBM with $m=4$, respectively. 
HBM can approximate sales data which BM cannot do.

Finally, we estimate the number of advertisements $B$. According to Eq.~(\ref{firstb}), it can be expressed by $B=Nb/a$.
The numbers calculated by parameters in Tab.~\ref{Tab:para} and ones derived by the least squares method are shown in Tab.~\ref{Tab:advs}.
From the result of HBM with $m=4$, we can understand that the number of advertisement influencing adopters for about 4 years from the product launch is about $10^4$.
On the other hand, the number from BM is 16.5. Owing to this, we can conclude that BM cannot approximate the iPod sales correctly.
Here, we note that the reason is not due to the peculiar method of fitting. If we estimate the number by using  parameters from the least squares method, it is 8.90 which is shown in the fourth column of Tab.~\ref{Tab:advs}: It becomes more and more unnatural.
\begin{table}
\caption{\label{Tab:advs}Number of advertisements, $B$, estimated from parameters $a$, $b$ and $N$ minimizing the product of SARE and SAE (the second the third column) and the residual sum of squares (the fourth column).}
\begin{center}
\begin{tabular*}{1.0\textwidth}{p{0.35\textwidth}p{0.19\textwidth}p{0.21\textwidth}p{0.12\textwidth}}
\hline
\hline
                                                       & BM                                &  HBM with $m=4$    & BM$^*$  \\
\hline
$a$                                                 & $1.42$                          & $4.50$                    & $1.95$   \\
$b$                                                 & $2.19\times10^{-10}$ & $0.000710$            & $2.08\times 10^{-7}$\\
$N \ [\times10^6]$                       & $1.07\times10^5$        & $66.4$                    & $83.4$ \\
\hline
$B$ \ (number of advertisements)  & $16.5$                          & $1.05\times 10^4$ & $8.90$ \\
\hline
\hline
\end{tabular*}
\end{center}
\end{table}

\section{Concluding remarks}

In this paper, we made it clear that BM is based on imitation among people and influence from advertisements, and 
proposed the new model, HBM, including a memory of how many adopters or advertisements a non-adopter met which BM lacks.
Additionally, we utilized the model to fit the iPod sales data, and so the better agreement was obtained than BM.
The iPod sales increased gradually until they peaked for the first time in  November 2005.
HBM is far more suitable for describing such a slow pace than BM, because all the non-adopters in our society did not purchase the iPod quickly after they met adopters or advertisements.

Other unsuitable points of approximating the iPod sales data by BM are the estimated total population $N$ and number of advertisements $B$: According to Tab.~\ref{Tab:para} and \ref{Tab:advs}, $N=1.07\times10^{11}$ and $B=16.5$.
However, there are strong doubts about whether the number of advertisements calculated by HBM, $B=1.05\times10^4$, is appropriate or not.
We are influenced by various types of advertisements, not only billboards which BM and HBM assume, but also advertisements on TV or Internet which always exist close to us. 
Therefore, BM and HBM are classical in terms of treating advertisements and it is necessary to model them practically.

\section*{Acknowledgments}
We would like to thank Suzuko Kawakami and members of astrophysics laboratory at Ochanomizu
University for extensive discussions.
This work was supported by JSPS Grant-in-Aid for Exploratory Research 12849926. 

\section*{Appendix}
\appendix
\setcounter{section}{1}

We discuss the symmetry and the constraint of BM by using the following nondimensionalized equation
\begin{equation}
\frac{{\rm d}p(\tau)}{{\rm d}\tau} = \left\{\frac{b}{a}+p(\tau)\right\}\left\{1-p(\tau)\right\} \ ,
\end{equation}
where $\tau\equiv ta$. This solution can be derived as 
\begin{equation}
p(\tau) = \frac{{\rm e}^{(1+b/a)\tau}-\frac{b}{a}\frac{1-p_0}{b/a+p_0}}
{{\rm e}^{(1+b/a)\tau}+\frac{1-p_0}{b/a+p_0}} \ ,
\label{eq:0}
\end{equation}
where $p_0\equiv p(0)$. Therefore,
\begin{equation}
\dot{p}(\tau) = \left(\frac{1+b/a}{{\rm e}^{(1+b/a)\tau}+\frac{1-p_0}{b/a+p_0}}\right)^2\frac{1-p_0}{b/a+p_0}{\rm e}^{(1+b/a)\tau} \ ,
\label{eq:d}
\end{equation}
and
\begin{equation}
\ddot{p}(\tau) = \left(\frac{1+b/a}{{\rm e}^{(1+b/a)\tau}+\frac{1-p_0}{b/a+p_0}}\right)^3\frac{1-p_0}{b/a+p_0}{\rm e}^{(1+b/a)\tau}\left(\frac{1-p_0}{b/a+p_0}-{\rm e}^{(1+b/a)\tau}\right) \ .
\label{eq:dd}
\end{equation}
Let us define $\tau$ where $\dot{p}(\tau)$ peaks as $\tau^*$, that is $\ddot{p}(\tau^*)=0$. According to Eq.~(\ref{eq:dd}), one can see the following relation satisfies:
\begin{equation}
{\rm e}^{(1+b/a)\tau^*} = \frac{1-p_0}{b/a+p_0} \ .
\label{eq:rel}
\end{equation}
Here, we introduce a deviation from $\tau^*$ by $\epsilon$, that is $\epsilon\equiv\tau-\tau^*$. By using the above relation, equation (\ref{eq:d}) becomes
\begin{equation}
\dot{p}(\tau) = \left(1+\frac{b}{a}\right)^2\left({\rm e}^{(1+b/a)\epsilon/2}+{\rm e}^{-(1+b/a)\epsilon/2}\right)^{-2}.
\label{re1}
\end{equation}
This is an even function of $\epsilon$, which means that $\dot{p}(\tau)$ is symmetric at the peak.

From Eq.~(\ref{eq:0}) with the relation (\ref{eq:rel}), we can obtain
\begin{equation}
p(\tau^*) = \frac{1}{2}\left(1-\frac{b}{a}\right) \ ,
\label{re2}
\end{equation}
from which we can see that $p(\tau^*)$ does not exceed $50\%$ because of $b/a\ge0$.

Note that theses results Eqs.~(\ref{re1}) and (\ref{re2}) are independent on the initial condition $p_0$.

\end{document}